\documentclass[aps, pra, preprint, superscriptaddress]{revtex4-2}

\usepackage{amssymb}
\usepackage{amsmath}
\usepackage{braket}
\usepackage{xfrac}
\usepackage{bbm}
\usepackage{xcolor}
\usepackage{tikz}
\usepackage{multirow}

\usepackage{hyperref}
\hypersetup{
    colorlinks=true,
    linkcolor=blue,
    filecolor=blue,      
    urlcolor=blue,
    citecolor = blue,
}

\newcommand{\tr}{\text{tr}}

\allowdisplaybreaks

\begin{document}
\nocite{*}
\title{Heisenberg Picture for Open Quantum Systems}
\author{Nachiket Karve}
\affiliation{Department of Physics, Indian Institute of Technology Kanpur, Kanpur 208016, Uttar Pradesh, India.}
\author{R. Loganayagam}
\affiliation{International Centre for Theoretical Sciences (ICTS-TIFR), Bangalore 560012, India.}

\begin{abstract}
In this note, we develop a framework to describe open quantum systems in the Heisenberg picture, i.e., via time evolving operator algebras. We point out the incompleteness of the 
previous proposals in this regard. We argue that a complete Heisenberg picture for an open quantum system involves multiple  \emph{image Heisenberg operators} for each system observable. For a given system observable, the number of such image operators is equal to the dimension of the environment Hilbert space. We derive a perturbative expression, accurate upto arbitrary orders in the system environment coupling, for these image operators in terms of a single \emph{one point operator}. This expression depends non-linearly on the state of the environment. This perturbative expression can equivalently be thought of as deforming the operator product on the Hilbert space of the open quantum system.  In the Markovian limit, the one point operator evolves by an adjoint Lindblad equation. We illustrate these ideas using  a simple spin system.
\end{abstract}

\maketitle

\section{Introduction}
Any quantum system which interacts with the environment is called an open quantum system. 
The aim of this note is to understand how such systems can be described in Heisenberg picture. The conventional treatments of open quantum systems proceed via Schrodinger picture
\cite{breuer2002theory}, i.e., one starts with the  full Hilbert space formed by the direct product of the Hilbert spaces of the system and that of the environment and then defines
the state of the system via a reduced density matrix obtained by  tracing out the environment degrees of freedom. The dynamics is then encoded in the evolution of such 
a reduced density matrix. The question we are asking is as follows : can one instead 
describe the same dynamics via an evolution of observables/operator algebra ? 

The answer to this question is non-trivial for the following reason : given a system
and an environment interacting with each other, consider a Heisenberg operator which starts off initially as a purely system operator (i.e., it is a tensor product of an operator acting
on the system Hilbert space and the identity operator on the environment Hilbert space).
Under Heisenberg evolution, such an operator generically evolves into an operator which 
is no more a tensor-product-- this is just the statement of entanglement stated in Heisenberg picture. It is hence unclear a priori how to project this evolution into
an evolution of a single system operator, the `reduced Heisenberg operator' so to speak. 

A minimal requirement of this projection is this : any $N$ point function of the 
Heisenberg evolved system operators computed in the total system should equivalently be computable through $N$ point function of such a `reduced Heisenberg operator'. As we will
argue below, it is impossible to satisfy this requirement : \emph{there is no single
system operator whose $N$ point functions will reproduce the $N$ point functions of Heisenberg evolved system operators}. We will however see that we can reproduce
the required $N$ point functions if we allow for multiple time-dependent system operators
(the number of such operators is equal to the dimension of the environment Hilbert space). 
We will call these the \textit{image operators} acting on the system Hilbert space.

While the above statement follows from a straightforward application of quantum mechanics, the implications are somewhat perplexing. Consider the example of a quantum particle coupled to an environment with an infinite dimensional Hilbert space. A direct application of the statement above implies that while there is a single position operator for this particle in the Schrodinger picture, one requires an infinite number of Heisenberg position operators
to fully describe all correlation functions. A natural question would be to ask how these
multiple image operators are related to each other ? How do they differ in their dynamics ?
One objective of this note is to answer these queries.

We have not found a discussion of the aforementioned conceptual difficulties and their resolution in the existing literature. There have been previous attempts to understand open quantum systems in the Heisenberg picture \cite{RevModPhys.89.015001, kheir2019open, Clark_2010, doi:10.1063/1.4884300}. Some attempts include calculating the radiative corrections to atomic transition frequencies \cite{PhysRevLett.30.456}, and studying the non-Markovian nature of bosonic and fermionic baths \cite{PhysRevA.87.012110}. Here the adjective `non-Markovian' denotes the fact that the bath/environment retains memory, viz.,
the environment  correlation functions do not decay more quickly than the system correlations.

When the correlations in the environment do decay quickly, one can write down a ``master equation" which determines the Schrodinger evolution of the reduced density matrix \cite{doi:10.1063/1.1731409, 10.1143/PTP.20.948}. These master equations have a 
Lindblad form \cite{lindblad, doi:10.1063/1.522979, KOSSAKOWSKI1972247}. It is natural
then to shift this evolution on to the system operators and obtain a  Heisenberg picture with operators  evolving via the adjoint master equation\cite{Breuer1998,breuer2002theory}. This approach identifies a single system operator that reproduces the correct one point function of a Heisenberg evolved system operator and focuses on the evolution of this operator. It should be noted however that this operator, which we will call the \emph{one point operator}, \underline{does not} reproduce the correct higher point functions.

As mentioned above, a full Heisenberg picture for an open quantum system requires the full set of image Heisenberg operators to reproduce all correlation functions correctly. The central aim of this note is to describe how these image Heisenberg operators are related to
the one point operator that appears in the discussion above. We will see that a precise
relation can be written down to arbitrary orders in system bath coupling. This relation however turns out to depend non-linearly on the state of the environment. As we will demonstrate later, an alternate way to interpret this relation is to think of it as a
deformation of operator products within the  Hilbert space of open quantum system.

The original motivation of this work is to study open quantum field theories,i.e., dissipative effective field theories obtained by integrating out fast degrees of freedom, fluid dynamics being the paradigmatic example in this regard. The Heisenberg picture is natural and convenient in this context. More generally, solving for the Schrodinger evolution of the full reduced density matrix  might often be a difficult endeavour whereas focusing on the Heisenberg evolution of a subset of observables of experimental interest could be tractable. This point of view already underlies the hierarchy of equations approach\cite{PhysRevA.75.052108}, where the evolution of $N$-point correlation functions is described in terms of higher point correlation functions -- these are the Schwinger-Dyson equations in the parlance of quantum field theory. Such equations for correlators can
be then derived from the operator formalism described here.

To summarise, our goal here  is to develop a general framework that can be used to compute correlation functions and expectation values of system observables in the simplest way possible. We would also like to describe and compute Heisenberg picture operators corresponding to these observables without making any assumptions about the nature of the system.  

We will conclude this introduction with an outline: in section \S\ref{sec_problem}, we systematically define a set of reduced Heisenberg operators corresponding to a particular observable, which we call the image operators. We then introduce and calculate the reduced $N$-point operators from the image operators, which help us calculate correlation functions. In section \S\ref{sec_soln}, we determine the time evolution of the image operators and the one point operator. In section \S\ref{sec_1ptop}, we show that this set of image operators corresponding to an observable can be written in terms of the one point operator in a way that depends non-linearly on the bath state. This gives an alternate prescription for the
computation of $N$-point functions in terms of a deformed product. In section \S\ref{sec_markovian}, we examine the Markovian limit, and get a Lindblad equation for 1-point operators. In section \S\ref{sec_structure}, we show that the number of computations required to calculate a correlation function can be further reduced. Lastly, in section \S\ref{sec_example}, we demonstrate the effectiveness of the techniques developed by solving a simple spin system.

\section{Setting up the Problem}
\label{sec_problem}
In any open system, we have a system of interest interacting with a large bath. Let the Hamiltonian of the system under study be $H_0$, and that of the bath be $H_B$. Further, we assume that they interact via the interaction Hamiltonian $\lambda H_I$. Here $\lambda$ is a dimensionless parameter which defines the strength of coupling. Our goal is to calculate the evolution of various system observables in the limit that $\lambda$ is very small. The Hamiltonian of the combined system is:  $H = H_0 + H_B + \lambda H_I$. This full system is closed, and does not interact with any external sources. Thus, any operator $A$ in the Hilbert space of the full system evolves according to the Heisenberg equation of motion:
\begin{equation}
\frac{d}{dt} O(t) = \frac{i}{\hbar} [H(t), O(t)],
\label{eq_von_neumann}
\end{equation}
where we have assumed that none of the operators involved have explicit time dependence.

Generally, a bath contains many more degrees of freedom than the system under study. Solving the above equations involves keeping track of the bath as well. Since we are not interested in the evolution of the bath, we would like to reduce the problem to the system's Hilbert space. While doing so, we must make sure that the expectation values and correlation functions are preserved.

Let us assume that the interaction between the system and the bath was turned on at $t = 0$. Before turning on the interaction, the system and the bath were decoupled. Therefore, the density matrix of the full system can be written as: $\rho = \rho_0 \otimes \rho_B$, where $\rho_0$ is the system's density matrix, and $\rho_B$ is the bath's density matrix. Note that the density matrix does not evolve in the Heisenberg picture, and therefore we do not have to worry about the mixing of the system and bath density matrices.

The expectation value of any observable $O$ is calculated as:
\begin{align}
\langle O(t) \rangle &= \tr\{O(t)\rho_0 \otimes \rho_B\} = \tr_S\{\tr_B\{O(t)\rho_B\}\rho_0\} = \tr_S \{O_S(t) \rho_0\},
\end{align}
wherewe have defined $O_S(t) \equiv \tr_B\{O(t)\rho_B\}$. Note that the operator $O_S(t)$ is confined to the system's Hilbert space, since we have traced out the bath degrees of freedom. Further, we can calculate the expectation value of the observable $O$ by taking a trace of $O_S(t)\rho_0$ over the system. Thus, this operator can be thought of as the reduced Heisenberg operator corresponding to the observable $O$, as far as one point functions are concerned. We call this operator a 1-point operator. 

Similarly, we can define an $N$-point operator as:
\begin{equation}
(O_1(t_1)\dots O_N(t_N))_S = \tr_B\{O_1(t_1)\dots O_N(t_N)\rho_B\}.
\end{equation}
However, we run into problems while multiplying two reduced operators. We note that $(O_1(t_1)O_2(t_2))_S \neq O_{1S}(t_1)O_{2S}(t_2)$ due to the fact that the trace does not distribute over multiplication. Thus, the information about the operator $(O_1(t_1)O_2(t_2))_S$ is not fully contained inside $O_{1S}(t_1)$ and $O_{2S}(t_2)$.
This substantiates our assertion in the introduction: the specification of $O_S(t)$ alone
is not a complete Heisenberg picture description of the dynamics since it misses the 
information about higher point correlations.

To be able to construct any general $N$-point reduced operators, we will now define another set of intermediate operators. Consider the Hilbert spaces $\mathcal{H}_0$ and $\mathcal{H}_B$ of the system and the bath respectively. Let $\{\ket{i}| \ i \in I\}$ be an orthonormal basis of $\mathcal{H}_0$ and $\{\ket{\alpha}| \ \alpha \in A\}$ be an orthonormal basis of $\mathcal{H}_B$. Thus, the full Hilbert space $\mathcal{H} = \mathcal{H}_0 \otimes \mathcal{H}_B$ is spanned by the basis $\{\ket{i\alpha}| \ i \in I, \alpha \in A\}$. 

Next, we define the projection operator $T_{\alpha}: \mathcal{H}_0 \rightarrow \mathcal{H}$ as $T_{\alpha} = \ket{i \alpha}\bra{i}$; where we have used the Einstein summation convention for summing over $i$. Note that this projection operator maps states in the system's Hilbert space $\mathcal{H}_0$ to states in the full Hilbert space $\mathcal{H}_0 \otimes \mathcal{H}_B$. These projection operators obey the identity:
\begin{equation}
T_{\alpha}T_{\alpha}^{\dagger} = \mathbbm{1}.
\label{eq_identity}
\end{equation}
Here, the symbol $ \mathbbm{1}$ denotes the identity operator in the full Hilbert space $\mathcal{H}_0 \otimes \mathcal{H}_B$.
Finally, we define the operator $O_{\alpha \beta}(t)$ as:
\begin{equation}
O_{\alpha \beta}(t) = T_{\alpha}^{\dagger} O(t) T_{\beta}.
\end{equation}
We call these operators as  the \emph{image operators} $O_{\alpha \beta}(t)$ of the observable $O$. These image operators act on the system Hilbert space and the number of such image operators is equal to the dimension of the bath/environment.

We can construct the 2-point image operators $(O_1(t_1)O_2(t_2))_{\alpha \beta}$ from the 1-point image operators $O_{1 \alpha \beta}(t_1)$ and $O_{2 \alpha \beta}(t_2)$ using \eqref{eq_identity}:
\begin{equation}
(O_1(t_1)O_2(t_2))_{\alpha \beta} = O_{1 \alpha \gamma}(t_1)O_{2 \gamma \beta}(t_2).
\end{equation}
Similarly one can construct all $N$-point image operators and from them the $N$-point reduced operators:
\begin{equation}
(O_1(t_1)\dots O_N(t_N))_{S} = (O_1(t_1)\dots O_N(t_N))_{\alpha \beta}\rho_{B\beta\alpha}.
\end{equation}

Inserting the identity \eqref{eq_identity} between operators in the Heisenberg equation \eqref{eq_von_neumann}, we obtain the reduced Heisenberg equation for the image operators:
\begin{equation}
\frac{d}{dt}O_{\alpha \beta}(t) = \frac{i}{\hbar}\Big\{H_{\alpha \gamma}(t)O_{\gamma \beta}(t) - O_{\alpha \gamma}(t) H_{\gamma \beta}(t)\Big\}.
\label{eq_red_von_neumann}
\end{equation}
Note that this is an exact equation for a general open quantum system. We have thus established that the image operators give a complete description of the open quantum 
system in the Heisenberg picture, allowing us to compute arbitrary higher point functions pf the system observables. The coupled evolution of the image operators can also be then described exactly  by the above equation, provided  the image Hamiltonians describing the open quantum system are known.

\section{Perturbative Solution for System Operators}
\label{sec_soln}
In this section, we will solve the reduced Heisenberg equation for a general system operator in a perturbative expansion in the system bath coupling.As is customary, we will do this 
by passing on to an interaction picture.

Before $t = 0$, the system and bath do not interact and the system operator is confined to the system's Hilbert space,i.e.,
\begin{equation}
O(0) = O \otimes \mathbbm{1}_{B},
\end{equation}
where $O$ is a Schrodinger picture operator and $\mathbbm{1}_{B}$ is the identity operator in the bath Hilbert space. The image operators corresponding to $O$ are then given by: $O_{\alpha\beta}(0) = O\ \delta_{\alpha\beta}$ at $t=0$. Once the interaction is turned on, these image operators evolve and become non-zero for $\alpha \neq \beta$.

From \eqref{eq_von_neumann}, it is easy to see that the Hamiltonian of the entire system, $H(t)$, does not evolve in time:
\begin{equation}
H(t) = H(0) = H_0 + H_B + \lambda H_I,
\end{equation}
where, $H_0$, $H_B$, and $H_I$ are  Schrodinger picture operators. Thus the reduced Heisenberg equation becomes:
\begin{equation}
\frac{d}{dt}O_{\alpha \beta}(t) = \frac{i}{\hbar}\Big\{H_{\alpha \gamma}O_{\gamma \beta}(t) - O_{\alpha \gamma}(t) H_{\gamma \beta}\Big\},
\end{equation}
where $H_{\alpha\beta}$s are  Schrodinger picture operators. 

The Hamiltonian of the system $H_0$ is confined to the Hilbert space $\mathcal{H}_0$ and hence the corresponding image operators are given by $H_{0\alpha \beta} = H_0\ \delta_{\alpha \beta}$. On the other hand, the bath Hamiltonian has the image operators $H_{B\alpha\beta} = H_{B\alpha\beta}\ \mathbbm{1}_0$ with $\mathbbm{1}_0$ being the identity operator in $\mathcal{H}_0$ and $H_{B\alpha\beta}$ being the $(\alpha, \beta)^{th}$ element of $H_B$. Further, for simplicity, let us assume that the basis $\{\ket{\alpha}\}$ of the Hilbert space $\mathcal{H}_B$ is the energy eigen-basis of $H_B$ with $H_B\ket{\alpha} = E_{\alpha}\ket{\alpha}$.  Then the reduced Heisenberg equation takes the form
\begin{align}
\frac{d}{dt}O_{\alpha \beta}(t) =& \frac{i}{\hbar}[H_0, O_{\alpha\beta}(t)] + \frac{i}{\hbar}(E_{\alpha}-E_{\beta})O_{\alpha\beta}(t)+ \frac{i\lambda}{\hbar}\Big\{H_{I\alpha \gamma}O_{\gamma \beta}(t) - O_{\alpha \gamma}(t) H_{I\gamma \beta}\Big\}.
\end{align}

We can pass on to the interaction picture by extending the standard steps to image operators. We define the interaction picture operators $\tilde{O}_{\alpha\beta}(t)$ and $\tilde{H}_{I\alpha\beta}(t)$ as:
\begin{subequations}
\begin{equation}
\tilde{O}_{\alpha\beta}(t) \equiv U_0(t) \ O_{\alpha\beta}(t) \ U_0^{\dagger}(t) \ e^{-\tfrac{i(E_{\alpha}-E_{\beta})t}{\hbar}},
\label{eq_def_t_o}
\end{equation}
\begin{equation}
\tilde{H}_{I\alpha\beta}(t) \equiv U_0(t) \ H_{I\alpha\beta} \ U_0^{\dagger}(t) \ e^{-\tfrac{i(E_{\alpha}-E_{\beta})t}{\hbar}},
\label{eq_def_t_h_i}
\end{equation}
\end{subequations}
where $U_0(t) = e^{-\tfrac{iH_0 t}{\hbar}}$ is the system evolution operator before environmental interactions are turned on. Inserting these into the above equation, we get:
\begin{equation}
\frac{d}{dt}\tilde{O}_{\alpha \beta}(t) = \frac{i\lambda}{\hbar}\Big\{\tilde{H}_{I\alpha \gamma}(t)\tilde{O}_{\gamma \beta}(t) - \tilde{O}_{\alpha \gamma}(t) \tilde{H}_{I\gamma \beta}(t)\Big\}.
\label{eq_pert}
\end{equation}
This is the reduced Heisenberg equation in the interaction picture. This equation can now be solved perturbatively in $\lambda$. 

Up to first order in $\lambda$, we obtain
\begin{equation}
\tilde{O}_{\alpha\beta}(t) = O\ \delta_{\alpha\beta} + \frac{i\lambda}{\hbar}\int_0^t dt_1 [\tilde{H}_{I\alpha\beta}(t_1), O] + \mathcal{O}(\lambda^2).
\end{equation}
A solution to arbitrary order in $\lambda$ can be obtained via  the Dyson series:
\begin{equation}
\tilde{O}_{\alpha\beta}(t) = \tilde{U}^{\dagger}_{I\gamma\alpha}(t)\ O\ \tilde{U}_{I\gamma\beta}(t),
\label{eq_toab_gen}
\end{equation}
where the evolution operator $\tilde{U}_{I\alpha\beta}(t)$ has the perturbative expansion
\begin{equation}
\tilde{U}_{I\alpha\beta}(t) = \sum_{n=0}^{\infty} \Big(\frac{-i\lambda}{\hbar}\Big)^n 
\ \tilde{K}^{(n)}_{\alpha\beta}(t).
\end{equation}
Here, $\tilde{K}^{(n)}$ is the time-ordered product of interaction Hamiltonians, as is familar from time-dependent perturbation theory:
\begin{equation}
\tilde{K}^{(n)}_{\alpha\beta}(t) = \int_0^{t}dt_1\dots\int_0^{t_{n-1}}dt_n \ \tilde{H}_{I\alpha\gamma_1}(t_1)\dots\tilde{H}_{I\gamma_{n-1}\beta}(t_n),
\end{equation}
In particular, we have $\tilde{K}^{(0)}_{\alpha\beta}(t) = \delta_{\alpha\beta}$. We can now 
revert back to the Heisenberg picture to write down  a perturbative expansion for the image operators
\begin{align}
O_{\alpha\beta}(t) &= U^{\dagger}_{I\gamma\alpha}(t)\ U_0^{\dagger}(t)\ O\ U_0(t)\ U_{I\gamma\beta}(t)\notag\\&
= \sum_{n=0}^{\infty}\Big(\frac{\lambda}{\hbar}\Big)^n\sum_{r=0}^{n}i^{n-2r} \ K^{(n-r)\dagger}_{\gamma\alpha}(t)U_0^{\dagger}(t)OU_0(t) K^{(r)}_{\gamma\beta}(t),
\label{eq_oab_gen}
\end{align}
where we have defined
\begin{equation}
K^{(n)}_{\alpha\beta}(t) \equiv e^{\frac{i(E_{\alpha}-E_{\beta})t}{\hbar}}U_0^{\dagger}(t)\tilde{K}^{(n)}_{\alpha\beta}(t)U_0(t).
\end{equation}

In order to avoid clutter in our subsequent expressions, we define the super-operators $P^{(n)}_{\alpha\beta}$, and $P_S^{(n)}$ acting on a system operator $A(t)$ as:
\begin{equation}
P^{(n)}_{\alpha\beta}A(t) \equiv \sum_{r=0}^n i^{n-2r}K^{(n-r)\dagger}_{\gamma\alpha}(t)A(t)K^{(r)}_{\gamma\beta}(t),
\label{eq_pab}
\end{equation}
and
\begin{equation}
P^{(n)}_S A(t) \equiv P^{(n)}_{\alpha\beta}A(t) \rho_{\beta\alpha}.
\label{eq_ps}
\end{equation}
Contracting $O_{\alpha\beta}$ with the bath density matrix $\rho_{B\beta\alpha}$ gives a compact expression for  $O_S(t)$ in terms of these super-operators:
\begin{align}
O_S(t) =& \sum_{n=0}^{\infty}\Big(\frac{\lambda}{\hbar}\Big)^n P^{(n)}_SU_0^\dagger(t)OU_0(t).
\label{eq_ost_gen}
\end{align}

\section{The Deformed Operator Product : Image Operators In Terms of One Point Operators}
\label{sec_1ptop}
In this section, we show that all $N$-point operators can be expressed solely in terms of one point operators. We will then argue that these expressions can equivalently be interpreted as deforming the operator product in the Hilbert space of the open quantum system.

Using the super operators defined in the last section, the required expressions can be 
obtained as follows. First, we  rewrite the equation \eqref{eq_ost_gen} in the form
\begin{align}
O_S(t) =& \ \Big\{ 1 + \sum_{n=1}^{\infty}\Big(\frac{\lambda}{\hbar}\Big)^nP^{(n)}_S\Big\}\ U_0^\dagger(t)\ O\ U_0(t),
\end{align}
and then invert it via multinomial expansion to give
\begin{equation}
\begin{split}
U_0^\dagger(t)\ O\ U_0(t) &= \ \Big\{ 1 + \sum_{n=1}^{\infty}\Big(\frac{\lambda}{\hbar}\Big)^nP^{(n)}_S\Big\}^{-1} O_S(t)\\
&=  \sum_{k=0}^{\infty} \sum_{n_1, \dots, n_k = 1}^\infty (-1)^k\left(\frac{\lambda}{\hbar}\right)^{n_1+\dots + n_k} P^{(n_1)}_S \dots P^{(n_k)}_S O_S(t).
\end{split}
\end{equation}
In the second step, we can  use the equation \eqref{eq_oab_gen} to express $O_{\alpha\beta}(t)$ in terms of $O_S(t)$:
\begin{align}
O_{\alpha\beta}(t) = \sum_{n=0}^\infty\sum_{k=0}^{\infty} \sum_{n_1, \dots, n_k = 1}^\infty (-1)^k\left(\frac{\lambda}{\hbar}\right)^{n+n_1+\dots +n_k} P^{(n)}_{\alpha\beta}P^{(n_1)}_S \dots P^{(n_k)}_S O_S(t).
\label{eq_oab_ost}
\end{align}
This is the central result of this note. As advertised, the above expression shows that all the image operators can be expressed in terms of one point operators. 

As all $N$ point operators can be reconstructed using image operators, it follows that reduced $N$ point operators can also be written in terms of the one point operators. However, as we noted before,  the reduced $N$ point operators are not simple products of 
the one point operator. The one point operators have to be dressed into the image operators
using the above formula and then multiplied at the level of image operators with appropriate 
index contractions. This procedure can be summarised by saying that the rule for operator
product on the system Hilbert space has to be modified.

Let us define a new operator product on the Hilbert space of an open quantum state such that 
\begin{equation}\label{eq_OP}
\begin{split}
O_S(t_1) \star O_S(t_2) \ldots \star O_S(t_N) \equiv O_{\alpha_1\alpha_2}(t_1)\ O_{\alpha_2\alpha_3}(t_2)\ldots O_{\alpha_N\alpha_{N+1}}(t_N)\ \rho_{B\alpha_{N+1}\alpha_N}\ .
\end{split}
\end{equation}
The way to interpret this expression is to think of the RHS using the equation \eqref{eq_oab_ost} which then defines a deformed operator product on the system Hilbert space. This shows that the new product depends non-trivially on the state of the bath. Once such a deformed operator product is defined, we can, in principle, remove all references to the image operators at the cost of working with a complicated product.

One can further write down an equation that expresses the evolution of all $N$-point operators in terms of 1-point operators, that is local in time. For example, $\frac{d}{dt}O_S(t)$ can be expressed as:
\begin{align}
\frac{d}{dt}O_S(t) =& \frac{i}{\hbar}[H_0, O_S(t)] \notag\\& + \sum_{n=1}^\infty\sum_{k=0}^{\infty} \sum_{n_1, \dots, n_k=1}^{\infty}(-1)^k \Big(\frac{\lambda}{\hbar}\Big)^{n+n_1+\dots +n_k} \mathcal{D}_t P^{(n)}_{S}P^{(n_1)}_S \dots P^{(n_k)}_S O_S(t),
\label{eq_evolution_1po}
\end{align}
where, the super-operator $\mathcal{D}_tP^{(n)}_S$ acting on an operator $A(t)$ is defined as:
\begin{align}
\mathcal{D}_tP^{(n)}_S A(t) =& \sum_{r=0}^n i^{n-2r}U_0^\dagger(t)\frac{d}{dt}\Big[U_0(t)K^{(n-r)\dagger}_{\gamma\alpha}(t)U_0^\dagger(t)\Big]U_0(t)A(t)K^{(r)}_{\gamma\beta}(t) \rho_{\beta\alpha} \notag\\&
+ \sum_{r=0}^n i^{n-2r}K^{(n-r)\dagger}_{\gamma\alpha}(t)A(t)U_0^\dagger(t)\frac{d}{dt}\Big[U_0(t)K^{(r)}_{\gamma\beta}(t)U_0^\dagger(t)\Big]U_0(t) \rho_{\beta\alpha}.
\end{align}

\section{Markovian Limit}
\label{sec_markovian}
A Markovian system is one in which the system does not remember its previous state. In the Schordinger picture, the density matrix obeys the Lindblad equation in the Markovian limit \cite{lindblad, doi:10.1063/1.522979, KOSSAKOWSKI1972247}. Under certain approximations, the reduced one point operator obeys a similar equation, called the adjoint master equation \cite{breuer2002theory}. We would like to understand which assumptions lead to a Lindblad equation in our formalism.

A general interaction Hamiltonian can be written as:
\begin{equation}
H_I = \sum_i R^i \otimes S^i,
\end{equation}

where $R^i$ is an operator in the system's Hilbert space, and $S^i$ is an operator in the bath's Hilbert space. Thus, the corresponding image operators of $H_I$ are:
\begin{equation}
H_{I\alpha\beta} = \sum_i R^i S^i_{\alpha\beta}.
\end{equation}

Note that since $S^i$ is confined to $\mathcal{H}_B$, $S^i_{\alpha\beta}$ is the $\alpha$, $\beta$ th element of $S^i$. From \eqref{eq_def_t_h_i}, we see that:
\begin{equation}
\tilde{H}_{I\alpha\beta}(t) =  \sum_i \tilde{R}^i (t) \tilde{S}^i_{\alpha\beta}(t),
\end{equation}

where, $\tilde{R}^i(t) = U_0(t)R^iU_0^\dagger(t)$, and $\tilde{S}^i_{\alpha\beta}(t) = {S}^i_{\alpha\beta} \ e^{-\frac{i(E_{\alpha}-E_{\beta})t}{\hbar}}$. We claim that $O_S(t)$ obeys a Lindblad equation under the following assumptions:
\begin{itemize}
\item The first moment of the interaction Hamiltonian is zero. That is, $\tr_B\{\tilde{H}_I(t)\rho_B\} = 0$.
\item The two point bath correlation functions $\langle \tilde{S}^i(t)\tilde{S}^j(t-\tau)\rangle_B$ decay rapidly with $\tau$.
\item The two point bath correlation functions $\langle \tilde{S}^i(t)\tilde{S}^j(t-\tau)\rangle_B$ are independent of $t$.
\end{itemize}

The first assumption implies that:
\begin{equation}
\tilde{S}^i_{\alpha\beta}(t)\rho_{B\beta\alpha} = 0.
\label{eq_1st_mom}
\end{equation}

Differentiating this with respect to $t$, we get:
\begin{equation}
\sum_{\alpha\beta}(E_{\alpha}-E_{\beta})\tilde{S}^i_{\alpha\beta}(t)\rho_{B\beta\alpha} = 0.
\label{eq_1st_mom_2}
\end{equation}

Similarly, the third assumption implies that:
\begin{equation}
\sum_{\alpha\beta}(E_{\alpha}-E_{\beta})\tilde{S}^i_{\alpha\gamma}(t)\tilde{S}^i_{\gamma\beta}(t-\tau)\rho_{B\beta\alpha} = 0.
\label{eq_2nd_mom}
\end{equation}

Using these assumptions, equation \eqref{eq_evolution_1po} simplifies to:
\begin{align}
\frac{d}{dt}O_S(t) =& \ \frac{i}{\hbar}H_0O_S(t) + \Big(\frac{i\lambda}{\hbar}\Big)^2\sum_{\omega, \omega'}
\sum_{ij} J^{ij}(\omega) \{A^{i\dagger}_{\omega}A^j_{\omega'}O_S(t) - A^{i\dagger}_{\omega}O_S(t)A^j_{\omega'}\} + \text{h.c.} + \mathcal{O}(\lambda^3),
\end{align}

where,
\begin{equation}
J^{ij}(\omega) = \int_0^{\infty}d\tau \ e^{-i\omega\tau}\tilde{S}^i_{\alpha\gamma}(0)\tilde{S}^j_{\gamma\beta}(-\tau)\rho_{B\beta\alpha},
\end{equation}

and, $A^i$s are the Fourier coefficients of $\tilde{R}^i(t)$:
\begin{equation}
\tilde{R}^i(t) = \sum_{\omega} e^{i\omega t}A^i_\omega.
\end{equation}

The above equation is of the Lindblad form.

\section{Structure of N-Point Operators}
\label{sec_structure}
In section \S\ref{sec_1ptop}, we showed that $N$-point operators can be constructed from one point operators, which reduces the number of calculations one has to perform to calculate $N$-point correlation functions. However, one can further reduce the number of calculations required to compute an $N$-point operator. One finds that $N$-point operators have a very peculiar structure which can be exploited to reduce the complexity of any calculation.

Using the definitions \eqref{eq_pab} and \eqref{eq_ps}, we can expand out equation \eqref{eq_oab_ost}. We describe a systematic way of writing down the $n$th order term in this expansion: We partition $n$ into an even number of partitions $\{n_1, m_1, \dots, n_k, m_k\}$, such that:
\begin{subequations}
\begin{equation}
\sum_{i=1}^k (n_i + m_i) = n,
\end{equation}
\begin{equation}
n_1+m_1 \geq 0,
\label{eq_cond1}
\end{equation}
\begin{equation}
n_i + m_i > 0, \ i = 2, \dots, k.
\label{eq_cond2}
\end{equation}
\end{subequations}

We construct the term corresponding to this partition as follows: Begin with $n_k$ and $m_k$. We multiply $O_S(t)$ by the operator $\Big[\Big(\frac{-i\lambda}{\hbar}\Big)^{n_k}K^{(n_k)}_{\gamma_k\alpha_k}(t)\Big]^{\dagger}$ from the left, and by the operator $\Big[\Big(\frac{-i\lambda}{\hbar}\Big)^{m_k}K^{(m_k)}_{\gamma_k\beta_k}(t)\Big]$ from the right. Here, a sum over $\gamma_k$ is implied. We contract this with the bath density matrix $\rho_{B\beta_k\alpha_k}$. We do a similar operation for $n_{k-1}$ and $m_{k-1}$. We then continue in a similar fashion until $n_2$ and $m_2$. For $n_1$ and $m_1$, we perform the same operation, except that we do not contract with the bath density matrix this time. At the end, we add a factor of $(-1)^{k-1}$. Finally, we get the term:
\begin{equation}\label{eq_nth_order_form}
\begin{split}
&(-1)^{k-1}\Big[\Big(\frac{-i\lambda}{\hbar}\Big)^{n_1}K^{(n_1)}_{\gamma_1\alpha}(t)\Big]^{\dagger}
\Big[\Big(\frac{-i\lambda}{\hbar}\Big)^{n_2}K^{(n_2)}_{\gamma_2\alpha_2}(t)\Big]^{\dagger} \dots
\Big[\Big(\frac{-i\lambda}{\hbar}\Big)^{n_k}K^{(n_k)}_{\gamma_k\alpha_k}(t)\Big]^{\dagger}O_S(t)\\
&\Big[\Big(\frac{-i\lambda}{\hbar}\Big)^{m_k}K^{(m_k)}_{\gamma_k\beta_k}(t)\Big] \dots \Big[\Big(\frac{-i\lambda}{\hbar}\Big)^{m_2}K^{(m_2)}_{\gamma_2\beta_2}(t)\Big] \Big[\Big(\frac{-i\lambda}{\hbar}\Big)^{m_1}K^{(m_1)}_{\gamma_1\beta}(t)\Big]\rho_{B\beta_2\alpha_2}\dots\rho_{B\beta_k\alpha_k},
\end{split}
\end{equation}
corresponding to the partition $\{n_1, m_1, \dots, n_k, m_k\}$. We sum over all such terms corresponding to even partitions of $n$ to get the $n$th order term in the expansion of the image operator $O_{\alpha\beta}(t)$.

We can make a quick check of the validity of this method. When $O_{\alpha\beta}(t)$ is contracted with $\rho_{B\beta\alpha}$, we get $O_S(t)$ by definition. We expect the above method to give us the same result. Consider the two partitions of $n$: $\{n_1, m_1, \dots, n_k, m_k\}$, and $\{0, 0, n_1, m_1, \dots, n_k, m_k\}$, such that $n_1+m_1 > 0$. The terms in the expansion of $O_{\alpha\beta}(t)$ corresponding to these partitions can be calculated by using \eqref{eq_nth_order_form}. If we contract both these terms with $\rho_{B\beta\alpha}$, then they differ only by a negative sign. Thus, both of them cancel out each other. The only remaining term is the one corresponding to the partition $\{0, 0\}$. Contracting this with $\rho_{B\beta\alpha}$ gives us $O_S(t)$, as expected.

For an $N$ point operator, we define its irreducible part, $I[O_1(t_1), \dots, O_N(t_N)]$ as the sum of the terms which cannot be broken into simpler terms. Diagrammatically, we represent this irreducible part by:
\begin{equation}
I[O_1(t_1), \dots, O_N(t_N)] =
\begin{tikzpicture}[baseline={([yshift=-.8ex]current bounding box.center)}]
\filldraw[black, fill = gray!30] (0, 0) circle (0.9);
\node at (0, 0) {1, \dots, N};
\end{tikzpicture}.
\end{equation}

We claim that all time ordered $N$ point operators can be broken down into a sum of irreducible operators. We demonstrate this by analysing the two point and three point operators. The two point image operators can be constructed from the one point image operators as:
\begin{equation}
(O_{1}(t_1)O_{2}(t_2))_{\alpha\beta} = O_{1\alpha\gamma}(t_1)O_{2\gamma\beta}(t_2).
\end{equation}

Thus, the $n$th order term in the two point image operator is formed by the $2k_1+2k_2$ partition of $n$. We will denote this partition by $\{n^{(1)}_1, m^{(1)}_1, \dots, n^{(1)}_{k_1}, m^{(1)}_{k_1}; n^{(2)}_1, m^{(2)}_1, \dots, n^{(2)}_{k_2}, m^{(2)}_{k_2}\}$. As before, this partition must satisfy the conditions \eqref{eq_cond1} and \eqref{eq_cond2}. We contract each partition with $\rho_{B\beta\alpha}$ to get the expansion of the two point operator, $(O_1(t_1)O_2(t_2))_{S}$. Table \ref{table_2pt} lists all possible non-zero partitions and their corresponding diagrams. All other partitions cancel out each other. This gives us the relation:
\begin{equation}\label{eq_2pt}
(O_1(t_1)O_2(t_2))_{S} = O_{1S}(t_1)O_{2S}(t_2) + I[O_{1S}(t_1), O_{2S}(t_2)].
\end{equation}

Note that this means that the irreducible part of the two point operator is nothing but the second order cumulant:
\begin{equation}
I[O_{1S}(t_1), O_{2S}(t_2)] = (O_1(t_1)O_2(t_2))_{S} - O_{1S}(t_1)O_{2S}(t_2).
\end{equation}

\begin{table}[h]
\centering
\begin{tabular}{|c|c|c|}
\hline
Partition & Term & Diagram \\ \hline
$n^{(1)}_i = m^{(1)}_i = n^{(2)}_i = m^{(2)}_i = 0$ & $O_{1S}(t_1)O_{2S}(t_2)$ & \begin{tikzpicture}[baseline={([yshift=-.8ex]current bounding box.center)}]
\filldraw[black, fill = gray!30] (0, 0) circle (0.3);
\filldraw[black, fill = gray!30] (1, 0) circle (0.3);
\draw (0.3, 0) -- (0.7, 0);
\node at (0, 0) {1};
\node at (1, 0) {2};
\end{tikzpicture} \\ \hline 
$n^{(1)}_1 + m^{(1)}_1 > 0$, $n^{(2)}_1 + m^{(2)}_1 > 0$ & $I[O_{1S}(t_1), O_{2S}(t_2)]$ & \begin{tikzpicture}[baseline={([yshift=-.8ex]current bounding box.center)}]
\filldraw[black, fill = gray!30] (0, 0) circle (0.5);
\node at (0, 0) {1, 2};
\end{tikzpicture} \\ \hline
\end{tabular}
\caption{Two Point Operator}
\label{table_2pt}
\end{table}

The three point image operators can be constructed from the one point image operators as:
\begin{equation}
(O_1(t_1)O_2(t_2)O_3(t_3))_{\alpha\beta} = O_{1\alpha\gamma}(t_1)O_{2\gamma\delta}(t_2)O_{3\delta\beta}(t_3).
\end{equation}

Again, as before, the $n$th order term in the three point image operator is formed by $2k_1+2k_2+2k_3$ partitions of $n$. They will be denoted by $\{n^{(1)}_1, m^{(1)}_1, \dots, n^{(1)}_{k_1}, m^{(1)}_{k_1}; n^{(2)}_1, m^{(2)}_1, \dots,\\ n^{(2)}_{k_2}, m^{(2)}_{k_2}; n^{(3)}_1, m^{(3)}_1, \dots, n^{(3)}_{k_3}, m^{(3)}_{k_3}\}$. To get the three point operator, $(O_1(t_1)O_2(t_2)O_3(t_3))_S$, we contract the three point image operators with $\rho_{B\beta\alpha}$. We list all partitions and their corresponding terms and diagrams in table \ref{table_3pt}.
\begin{table}[h]
\centering
\begin{tabular}{|c|c|c|}
\hline
Partition & Term & Diagram \\ \hline
$n^{(k)}_i = m^{(k)}_i = 0$, $k = 1, 2, 3$ & $O_{1S}(t_1)O_{2S}(t_2)O_{3S}(t_3)$ & \begin{tikzpicture}[baseline={([yshift=-.8ex]current bounding box.center)}]
\filldraw[black, fill = gray!30] (0, 0) circle (0.3);
\filldraw[black, fill = gray!30] (0.9, 0) circle (0.3);
\filldraw[black, fill = gray!30] (1.8, 0) circle (0.3);
\draw (0.3, 0) -- (0.6, 0);
\draw (1.2, 0) -- (1.5, 0);
\node at (0, 0) {1};
\node at (0.9, 0) {2};
\node at (1.8, 0) {3};
\end{tikzpicture} \\ \hline
\multirow{2}{2.1in}{
$n^{(1)}_1 + m^{(1)}_1 > 0$, $n^{(2)}_1 + m^{(2)}_1 > 0$, $n^{(3)}_i = m^{(3)}_i = 0$} & \multirow{2}{2.4in}{$\mathcal{W}_{1,2,3}\{O_{3S}(t_3)I[O_{1S}(t_1), O_{2S}(t_2)]\}$ }& \multirow{2}{0.7in}{\begin{tikzpicture}[baseline={([yshift=-.8ex]current bounding box.center)}]
\filldraw[black, fill = gray!30] (0, 0) circle (0.3);
\filldraw[black, fill = gray!30] (1, 0) circle (0.5);
\draw (0.3, 0) -- (0.5, 0);
\node at (0, 0) {3};
\node at (1, 0) {1, 2};
\end{tikzpicture}} \\ & & \\ \hline
\multirow{2}{2.1in}{
$n^{(1)}_1 + m^{(1)}_1 > 0$, $n^{(2)}_i = m^{(2)}_i = 0$, $n^{(3)}_1 + m^{(3)}_1 > 0$} & \multirow{2}{2.4in}{$\mathcal{W}_{1,2,3}\{O_{2S}(t_2)I[O_{3S}(t_3), O_{1S}(t_1)]\}$ }& \multirow{2}{0.7in}{\begin{tikzpicture}[baseline={([yshift=-.8ex]current bounding box.center)}]
\filldraw[black, fill = gray!30] (0, 0) circle (0.3);
\filldraw[black, fill = gray!30] (1, 0) circle (0.5);
\draw (0.3, 0) -- (0.5, 0);
\node at (0, 0) {2};
\node at (1, 0) {3, 1};
\end{tikzpicture}} \\ & & \\ \hline
\multirow{2}{2.1in}{
$n^{(1)}_i = m^{(1)}_i = 0$, $n^{(2)}_1 + m^{(2)}_1 > 0$, $n^{(3)}_1 + m^{(3)}_1 > 0$} & \multirow{2}{2.4in}{$\mathcal{W}_{1,2,3}\{O_{1S}(t_1)I[O_{2S}(t_2), O_{3S}(t_3)]\}$ }& \multirow{2}{0.7in}{\begin{tikzpicture}[baseline={([yshift=-.8ex]current bounding box.center)}]
\filldraw[black, fill = gray!30] (0, 0) circle (0.3);
\filldraw[black, fill = gray!30] (1, 0) circle (0.5);
\draw (0.3, 0) -- (0.5, 0);
\node at (0, 0) {1};
\node at (1, 0) {2, 3};
\end{tikzpicture}} \\ & & \\ \hline
\multirow{2}{2.1in}{
$n^{(1)}_1 + m^{(1)}_1 > 0$, $n^{(2)}_1 + m^{(2)}_1 > 0$, $n^{(3)}_1 + m^{(3)}_1 > 0$} & \multirow{2}{2in}{$I[O_{1S}(t_1), O_{2S}(t_2), O_{3S}(t_3)]$ }& \multirow{2}{0.5in}{\begin{tikzpicture}[baseline={([yshift=-.8ex]current bounding box.center)}]
\filldraw[black, fill = gray!30] (0, 0) circle (0.6);
\node at (0, 0) {1, 2, 3};
\end{tikzpicture}} \\ & & \\ \hline
\end{tabular}
\caption{Three Point Operator}
\label{table_3pt}
\end{table}

Note that we have introduced the operator $\mathcal{W}_{1,2,3}$. This operator makes sure that the operator product is ordered such that $O_1$ comes before $O_2$, and $O_2$ comes before $O_3$. For example, the third term in table \ref{table_3pt} is:
\begin{equation}
\mathcal{W}_{1,2,3}\{O_{2S}(t_2)I[O_{3S}(t_3), O_{1S}(t_1)]\} = (O_1(t_1)O_{2S}(t_2)O_3(t_3))_S - O_{1S}(t_1)O_{2S}(t_2)O_{3S}(t_3).
\end{equation}

Once again, we find that the irreducible part of the three point operator is the third order cumulant:
\begin{align}
I[O_{1S}(t_1) ,O_{2S}(t_2), O_{3S}(t_3)] =& \ (O_1(t_1)O_2(t_2)O_3(t_3))_S \notag\\& - [(O_{1}(t_1)O_{2}(t_2)O_{3S}(t_3))_S - O_{1S}(t_1)O_{2S}(t_2)O_{3S}(t_3)]
\notag\\& - [(O_{1}(t_1)O_{2S}(t_2)O_{3}(t_3))_S - O_{1S}(t_1)O_{2S}(t_2)O_{3S}(t_3)]
\notag\\& - [(O_{1S}(t_1)O_{2}(t_2)O_{3}(t_3))_S - O_{1S}(t_1)O_{2S}(t_2)O_{3S}(t_3)] \notag\\&
- O_{1S}(t_1)O_{2S}(t_2)O_{3S}(t_3).
\end{align}

\section{Example: Two Qubit System}
\label{sec_example}
We demonstrate the usefulness of the methods developed above by solving a simple system. Consider a system of two spin-half particles with spin operators $\overline{S}_1$ and $\overline{S}_2$ interacting via the Hamiltonian:
\begin{equation}
H = \lambda \ \overline{S}_1 \cdot \overline{S}_2.
\end{equation}

We consider the first particle to be our system and the second particle to be the bath. We will also assume that the bath has a density matrix of the form:
\begin{align}
\rho_B &= (1-c)\ket{\uparrow}\bra{\uparrow} + c \ket{\downarrow}\bra{\downarrow}
= \begin{pmatrix}
1-c & 0\\ 0 & c
\end{pmatrix}.
\end{align}

This system can be solved exactly, and hence it is easy to verify the results of our perturbative method.

The interaction Hamiltonian of the system is $H_I = \overline{S}_1 \cdot \overline{S}_2$. Thus, the image operators of $H_I$ are:
\begin{align}
&H_{I\uparrow \uparrow} = \frac{\hbar^2}{4}\begin{pmatrix}
1 & 0\\ 0 & -1
\end{pmatrix},
H_{I\uparrow \downarrow} = \frac{\hbar^2}{2}\begin{pmatrix}
0 & 0\\ 1 & 0
\end{pmatrix},
H_{I\downarrow \uparrow} = \frac{\hbar^2}{2}\begin{pmatrix}
0 & 1\\ 0 & 0
\end{pmatrix},
H_{I\downarrow \downarrow} = \frac{\hbar^2}{4}\begin{pmatrix}
-1 & 0\\ 0 & 1
\end{pmatrix}.
\end{align}

Using this, we can calculate $K^{(1)}_{\alpha\beta}(t)$ and $K^{(2)}_{\alpha\beta}(t)$:
\begin{align}
&K_{\uparrow \uparrow}^{(1)}(t) = \frac{\hbar^2t}{4}\begin{pmatrix}
1 & 0\\ 0 & -1
\end{pmatrix},
K_{\uparrow \downarrow}^{(1)}(t) = \frac{\hbar^2t}{2}\begin{pmatrix}
0 & 0\\ 1 & 0
\end{pmatrix},\notag\\&
K_{\downarrow \uparrow}^{(1)}(t) = \frac{\hbar^2t}{2}\begin{pmatrix}
0 & 1\\ 0 & 0
\end{pmatrix},
K_{\downarrow \downarrow}^{(1)}(t) = \frac{\hbar^2t}{4}\begin{pmatrix}
-1 & 0\\ 0 & 1
\end{pmatrix},
\end{align}
\begin{align}
&K_{\uparrow \uparrow}^{(2)}(t) = \frac{\hbar^4t^2}{32}\begin{pmatrix}
1 & 0\\ 0 & 5
\end{pmatrix},
K_{\uparrow \downarrow}^{(2)}(t) = \frac{\hbar^4t^2}{8}\begin{pmatrix}
0 & 0\\ -1 & 0
\end{pmatrix}, \notag\\&
K_{\downarrow \uparrow}^{(2)}(t) = \frac{\hbar^4t^2}{8}\begin{pmatrix}
0 & -1\\ 0 & 0
\end{pmatrix},
K_{\downarrow \downarrow}^{(2)}(t) = \frac{\hbar^4t^2}{32}\begin{pmatrix}
5 & 0\\ 0 & 1
\end{pmatrix}.
\end{align}

We can contract $K^{(1)}_{\alpha\beta}(t)$ and $K^{(2)}_{\alpha\beta}(t)$ with the density matrix $\rho_{B\beta\alpha}$ to obtain $K^{(1)}_S(t)$ and $K^{(2)}_S(t)$:
\begin{subequations}
\begin{equation}
K^{(1)}_S(t) = (1-2c)\frac{\hbar^2t}{4}\begin{pmatrix}
1 & 0 \\ 0 & -1
\end{pmatrix},
\end{equation}
\begin{equation}
K^{(2)}_S(t) = \frac{\hbar^4t^2}{32}\begin{pmatrix}
1+4c & 0 \\ 0 & 5-4c
\end{pmatrix}.
\end{equation}
\end{subequations}

Let us now use our method to calculate the operator $S_{1xS}(t)$, which is the $x$-component of the first particle's spin. We will be using equation \eqref{eq_ost_gen}. Up to second order, we have:
\begin{align}
S_{1xS}(t) =& \ S_{1x} + \frac{i\lambda}{\hbar}[K^{(1)}_S(t), S_{1x}] + \Big(\frac{i\lambda}{\hbar}\Big)^2 \Big\{K^{(2)\dagger}_S(t)S_{1x} + S_{1x}K^{(2)}_S(t)\notag\\& - K^{(1)\dagger}_{\gamma\alpha}(t)S_{1x}K^{(1)}_{\gamma\beta}(t)\rho_{B\beta\gamma}\Big\} + \mathcal{O}(\lambda^3)\notag\\
=& \frac{\hbar}{2}\begin{pmatrix}
0 & 1 + \frac{1}{2}(1-2c)i\lambda\hbar t -\frac{1}{4}(\lambda\hbar t)^2 \\
1 - \frac{1}{2}(1-2c)i\lambda\hbar t -\frac{1}{4}(\lambda\hbar t)^2 & 0
\end{pmatrix} + \mathcal{O}(\lambda^3). 
\label{eq_s1x_2nd_order}
\end{align}

If we calculate $S_{1xS}(t)$ exactly, we find that:
\begin{equation}
S_{1xS}(t) = \frac{\hbar}{2}\begin{pmatrix}
0 & \frac{1}{2}[1+\cos(\lambda\hbar t)+i(1-2c)\sin(\lambda\hbar t)]\\
\frac{1}{2}[1+\cos(\lambda\hbar t)-i(1-2c)\sin(\lambda\hbar t)] & 0
\end{pmatrix},
\end{equation}

which agrees with our perturbative calculations.

Now that we have calculated the one point operator $S_{1xS}$ up to second order, we immediately have the two point operator up to first order. Note that equation \eqref{eq_2pt} can be written as:
\begin{equation}
(S_{1x}(t_1)S_{1x}(t_2))_S = S_{1xS}(t_1)S_{1xS}(t_2) + \mathcal{O}(\lambda^2).
\end{equation}

We can directly use the result \eqref{eq_s1x_2nd_order} and obtain:
\begin{align}
(S_{1x}(t_1)S_{1x}(t_2))_S =& \frac{\hbar}{2}\begin{pmatrix}
1 + \frac{1}{2}i(1-2c)(t_1-t_2)\lambda & 0 \\
0 & 1 - \frac{1}{2}i(1-2c)(t_1-t_2)\lambda
\end{pmatrix} 
+ \mathcal{O}(\lambda^2).
\end{align}

We can immediately verify that this agrees with the exact two point operator.

\section{Conclusion}

In this work, we have developed a technique to handle open quantum systems in Heisenberg picture. We started by defining $N$-point operators and image operators for system observables. We then perturbatively solved the reduced Heisenberg equation, and figured out the $N$-point operators and the image operators as a function of time. We then showed that all $N$-point operators can be expressed solely in terms of 1-point operators in a way that is dependent on the bath state. This expression can then be used to deform the  operator product on the Hilbert space of the open quantum system.

The Heisenberg picture described in this note is  novel in the way it associates multiple 
system operators to a single system observable. It is natural to speculate that these
multiplicity of operators play a crucial role in say describing decoherence in terms of observables rather than states. It is also clear that these image operators and their
inter-relations should encode the entanglement of the system with its environment. It would
be nice to make these statements precise. We also hope that the description in terms of the Heisenberg picture throws new light on discussions about interpretations of quantum mechanics. In particular, it would be interesting to explore the relation, if any, between the description of the open system in terms of multiple Heisenberg image operators and the 
Everett's many-world interpretation\cite{RevModPhys.29.454,manyworld}.

As we mentioned in the introduction, a main motivation of this work is to develop a systematic framework for open quantum field theories. The formalism described here can 
readily be adopted to field theories, i.e., quantum systems with infinite number of degrees of freedom. The Heisenberg evolution in QFTs famously involve UV divergences which need to be regulated and counter-termed in order to study renormalisation within these theories.
A related question is about operator products within open quantum field theories using the deformed product introduced in this work. We hope  to report on these issues in future.

\section*{Acknowledgements}
We would to thank Nilay Kundu and Mukund Rangamani for their valuable comments regarding this work. RL would like to acknowledge his debt to the people of India for their sustained and generous support to research in the basic sciences.

\bibliographystyle{plainnat}
\bibliography{refHeisOpen}

\end{document}